\documentclass[aps,showpacs,preprintnumbers,amsmath, amssymb]{revtex4}

\usepackage{float}
\usepackage{graphics,epsfig}
\usepackage{graphicx}
\usepackage{dcolumn}
\usepackage{bm}

\begin{document}
\baselineskip=0.8 cm

\title{{\bf Holographic superconductors in the AdS black-hole spacetime with a global monopole}}
\author{Songbai Chen}
\email{csb3752@163.com} \affiliation{ Institute of Physics and
Department of Physics, Hunan Normal University,  Changsha, Hunan
410081, P. R. China \\ Key Laboratory of Low Dimensional Quantum
Structures \\ and Quantum Control of Ministry of Education, Hunan
Normal University, Changsha, Hunan 410081, P. R. China}

\author{Liancheng Wang}
 \affiliation{ Institute of Physics and
Department of Physics, Hunan Normal University,  Changsha, Hunan
410081, P. R. China \\ Key Laboratory of Low Dimensional Quantum
Structures \\ and Quantum Control of Ministry of Education, Hunan
Normal University, Changsha, Hunan 410081, P. R. China}

\author{Chikun Ding}
 \affiliation{ Institute of Physics and
Department of Physics, Hunan Normal University,  Changsha, Hunan
410081, P. R. China \\ Key Laboratory of Low Dimensional Quantum
Structures \\ and Quantum Control of Ministry of Education, Hunan
Normal University, Changsha, Hunan 410081, P. R. China}

\author{Jiliang Jing}
\email{jljing@hunnu.edu.cn}
 \affiliation{ Institute of Physics and
Department of Physics, Hunan Normal University,  Changsha, Hunan
410081, P. R. China \\ Key Laboratory of Low Dimensional Quantum
Structures \\ and Quantum Control of Ministry of Education, Hunan
Normal University, Changsha, Hunan 410081, P. R. China}

\vspace*{0.2cm}
\begin{abstract}
\baselineskip=0.6 cm
\begin{center}
{\bf Abstract}
\end{center}

We study holographic superconductors in the Schwarzschild-AdS black
hole with a global monopole through a charged complex scalar field.
We calculate the condensates of the charged operators in the dual
conformal field theories (CFTs) and discuss the effects of the
global monopole on the condensation formation. Moreover, we compute
the electric conductive using the probe approximation and find that
the properties of the conductive are quite similar to those in the
Schwarzschild-AdS black hole. These results can help us know more
about holographic superconductors in the asymptotic AdS black holes.

\end{abstract}

\pacs{11.25.Tq,  04.70.Bw, 74.20.-z} \maketitle
\newpage
\section{Introduction}

One of the most significant discovery in the string theory is the
AdS/CFT correspondence which proposed  by Maldacena \cite{ads1} in
the last century. It tells us that the gravity theory in a
$(d+1)$-dimensional AdS spacetime can be described by a conformal
field theory on the $d$-dimensional boundary. The AdS/CFT
correspondence is a powerful tool to understand the strongly coupled
gauge theories \cite{ads2,ads3,ads4}. Recently, it has been applied
extensively to investigate some condensed matter phenomena, such as
superconductivity \cite{Hs0,Hs01,Hs02,Hs03}, superfluid
\cite{Hsf0,Hsf01}, and so on.

Superconductivity is a common phenomenon occurring in certain
materials at very low temperatures. The most striking feature of the
phenomenon is that the material has exactly zero electrical
resistance for a direct current (DC) and the exclusion of the
interior magnetic field when it is in the superconducting state. The
first model for the holographic superconductors in the AdS
black-hole spacetime is proposed by Hartnoll, Herzog and Horowitz
\cite{Hs0}. They considered the classical instability of a black
hole in AdS spacetime against perturbation by a charged scalar field
and found that if the temperature $T$ of the black hole is below a
critical temperature $T_c$ the instability of the black hole
emerges, which implies that the original Schwarzschild-AdS black
hole should be replaced by a new black hole with the charged scalar
hair. According to the AdS/CFT correspondence, the emergence of the
hairy AdS black hole means the formation of a charged condensation
in the dual CFTs \cite{Hsa1}. Moreover, the expectation values of
the charged operators undergo the U(1) symmetry breaking and then
the superconductive is formed.

Hartnoll, Herzog and Horowitz \cite{Hs0} also calculated the
electrical conductive of the charged condensation and found that the
condensation has zero electrical DC resistance, which is the same as
that obtained in the Bardeen-Cooper-Schrieffer (BCS) theory
\cite{bcs}. This has triggered many people to study holographic
superconductors in the various theories of gravity
\cite{a0,a01,a1,a2,a3,a4,a401,a40,a41,a42,
a5,a6,a7,a8,a8d1,a81,a9,a10,a100, a11,a12,a13,a14,a15,a16s,a17}.
Gregory \textit{et al} \cite{a0,a401} considered the holographic
superconductors in the Einstein-Gauss-Bonnet gravity and found that
the higher curvature corrections make condensation harder to form.
Cai \textit{et al} \cite{a40,a41,a42} investigated the holographic
superconductor models in the Ho\v{r}ava-Lifshitz gravity. Recent
studies also showed that there exist holographic superconductors in
the string/M theory \cite{a5,a6,a7,a8,a8d1}. The properties of
holographic superconductors at the zero-temperature limit have also
been considered extensively \cite{a81,a9,a10,a100}. These results
can help us to understand more about the holographic superconductors
in the asymptotical AdS black holes.

In this paper, we will investigate the holographic superconductors
in planar AdS black hole with a global monopole. A global monopole
is one of the topological defects which could be formed during phase
transitions in the evolution of the early Universe. The metric of
the black hole with a global monopole was obtained by Barriola and
Vilenkin \cite{mono1}, which arises from the breaking of global
$SO(3)$ symmetry of a triplet scalar field in a Schwarzschild
background. The presence of the global monopole results in that the
black hole has different topological structure from that of the
Schwarzschild black hole. The physical properties of the black hole
with a global monopole have been studied extensively in recent years
\cite{mono2,mono3,mono4,mono5}. The main purpose in this paper is to
see how the global monopole affects the holographic superconductors
in this asymptotic AdS black hole.

This paper is organized as follows. In Sec. II, we present the
metric describing a planar Schwarzschild-AdS black hole with a
global monopole. In Sec. III, we give the basic equations and study
numerically holographic superconductors in the planar black hole
background with a global monopole. Our results show that the global
monopole presents the different effects on the different
condensations. In Sec. IV, we calculate the electrical conductivity
of the charged condensates by ignoring the backreaction of the
dynamical matter field on the spacetime metric. Finally, in the last
section we present our conclusions.

\section{The planar Schwarzschild-AdS black hole with a global monopole}

The simplest model that gives rise to the global monopole is
described by the Lagrangian
\begin{eqnarray}
\mathcal{L}_{Gm}=\frac{1}{2}\partial_{\mu}\Phi^{a}\partial^{\mu}\Phi^{*a}-\frac{\gamma}{4}(\Phi^{a}\Phi^{*a}-\eta^2)^2,\label{l1}
\end{eqnarray}
where $\Phi^{a}$ is a multiplet of scalar field, $\eta$ is the
energy scale of symmetry breaking  and $\gamma$ is a constant. The
metric ansatz describing a planar black hole with a global monopole
can be taken as
\begin{eqnarray}
ds^2=-f(r)dt^2+\frac{1}{f(r)}dr^2+r^2(dx^2+dy^2).\label{m1}
\end{eqnarray}
In this case the field configuration describing a monopole is
\begin{eqnarray}
\Phi^{1}=\eta h(r) e^{ix}\cos{y},\;\;\;\;\;\Phi^{2}=\eta h(r)
e^{ix}\sin{y}. \label{sg1}
\end{eqnarray}
Using the Lagrangian (\ref{l1}) and metric (\ref{m1}), we can
calculate the energy-momentum tensor
\begin{eqnarray}
T_{\mu\nu}=2\frac{\partial\mathcal{L}_{Gm}}{\partial
g^{\mu\nu}}-\mathcal{L}_{Gm}g_{\mu\nu}\;,
\end{eqnarray}
and then obtain the nonzero components of the energy-momentum tensor
\begin{eqnarray}
T^{t}_t&=&\frac{1}{2}\eta^2h'(r)^2f(r)+\eta^2\frac{h(r)^2}{r^2}+\frac{\gamma}{4}\bigg[\eta^2h(r)^2-\eta^2\bigg]^2,\\
T^{r}_r&=&-\frac{1}{2}\eta^2h'(r)^2f(r)+\eta^2\frac{h(r)^2}{r^2}+\frac{\gamma}{4}\bigg[\eta^2h(r)^2-\eta^2\bigg]^2,\\
T^{x}_x&=&T^{y}_y=\frac{1}{2}\eta^2h'(r)^2f(r)+\frac{\gamma}{4}\bigg[\eta^2h(r)^2-\eta^2\bigg]^2.
\end{eqnarray}
As in Ref. \cite{mono1}, we take an approximation $h(r)= 1$ outside
the core. It is reasonable because that $h(r)$ grows linearly when
$r<(\eta\sqrt{\gamma})^{-1}$ and approaches exponentially unity as
soon as $r>(\eta\sqrt{\gamma})^{-1}$.

In the planar AdS background, the Einstein equations are given by
\begin{eqnarray}
&&\frac{f(r)'}{r}+\frac{f(r)}{r^2}-\frac{3}{L}+\frac{8\pi\eta^2}{r^2}=0,\label{ai1}\\
&&\frac{f(r)''}{2}+\frac{f(r)'}{r}-\frac{3}{L}=0\label{ai2},
\end{eqnarray}
where $L$ is the radius of AdS. An exact solution of the equations
(\ref{ai1}) and (\ref{ai2}) is
\begin{eqnarray}
f(r)=\frac{r^2}{L^2}-\tilde{b}-\frac{2M}{r},\label{me4}
\end{eqnarray}
where the parameter $\tilde{b}=8\pi\eta^2$. This solution describes
a planar Schwarzschild-AdS black hole with a global monopole. As
$\tilde{b}$ tends to zero, the spacetime reduces to a four
dimensional Schwarzschild-AdS black hole. The Hawking temperature of
the black hole (\ref{me4}) is
\begin{eqnarray}
T_H=\frac{3r^2_H-\tilde{b}}{4\pi r_H},
\end{eqnarray}
where $r_H$ is the event horizon of the black hole and we set the
AdS radius is $L=1$.

\section{The condensate of charged operators}

In order to investigate the holographic superconductors in the
background of the planar Schwarzschild-AdS black hole with a global
monopole, we need  a condensate through a charged scalar field. Here
we adopt to the probe approximation and neglect the backreaction of
the charged scalar field on the background. It must be pointed out
that this charged scalar field is not the multiplet scalar field
describing the global monopole. It may be an interesting topic to
study the holographic superconductors of the scalar field which
gives rise to the global monopole. However, it is very difficult
because that for the scalar field with the form (\ref{sg1}) the
coupled equation (\ref{e2}) cannot be separable in the following
calculations. Thus, in this paper we only consider the condensate of
an external charged scalar field in the background of a black hole
with a global monopole.

As in Ref. \cite{Hs0}, let us consider a Maxwell field and a charged
complex scalar field. The Lagrangian can be expressed as \cite{Hs0}
\begin{eqnarray}
\mathcal{L}=-\frac{1}{4}F^{\mu\nu}F_{\mu\nu}-
|\nabla_{\mu}\psi-iqA_{\mu}\psi|^2-m^2\psi^2,
\end{eqnarray}
where $F_{\mu\nu}$ is electromagnetic tensor and $\psi$ is a charged
complex scalar field. Adopting to the ansatz
\begin{eqnarray}
A_{\mu}=(\phi(r),0,0,0),\;\;\;\;\psi=\psi(r),
\end{eqnarray}
we can obtain the equations of motion for the complex scalar field
$\psi$ and electrical scalar potential $\phi(r)$ in the background
of the Schwarzschild-AdS black hole with a global monopole
\begin{eqnarray}
\psi^{''}+(\frac{f'}{f}+\frac{2}{r})\psi'
+\frac{q^2\phi^2\psi}{f^2}-\frac{m^2\psi}{f}=0,\label{e1}
\end{eqnarray}
and
\begin{eqnarray}
\phi^{''}+\frac{2}{r}\phi'
-\frac{2\psi^2}{f}\phi=0,\label{e2}
\end{eqnarray}
respectively. Here a prime denotes the derivative with respect to
$r$. Obviously, it is very difficult to obtain the nontrivial
analytical solutions to the nonlinear equations (\ref{e1}) and
(\ref{e2}). So we have to resort numerical method to solve above
equations. In general, the boundary condition on the scalar
potential $\phi$ near the black hole horizon $r\sim r_H$ is imposed
as $\phi=0$ so that its finite norm can be satisfied. Combining with
Eq. (\ref{e1}), it is easy to obtain that
$\psi=-\frac{3r^2_H-\tilde{b}}{2r_H}\psi'$, which means that the
complex scalar field $\psi$ is regular at $r=r_H$. At the spatial
infinite $r\rightarrow\infty$, the scalar field $\psi$ and the
scalar potential $\phi$ can be approximated as
\begin{eqnarray}
\psi=\frac{\psi^{(1)}}{r}+\frac{\psi^{(2)}}{r^2}+...,\label{b1}
\end{eqnarray}
and
\begin{eqnarray}
\phi=\mu-\frac{\rho}{r}+...\;.
\end{eqnarray}
\begin{figure}[ht]
\begin{center}
\includegraphics[width=6.8cm]{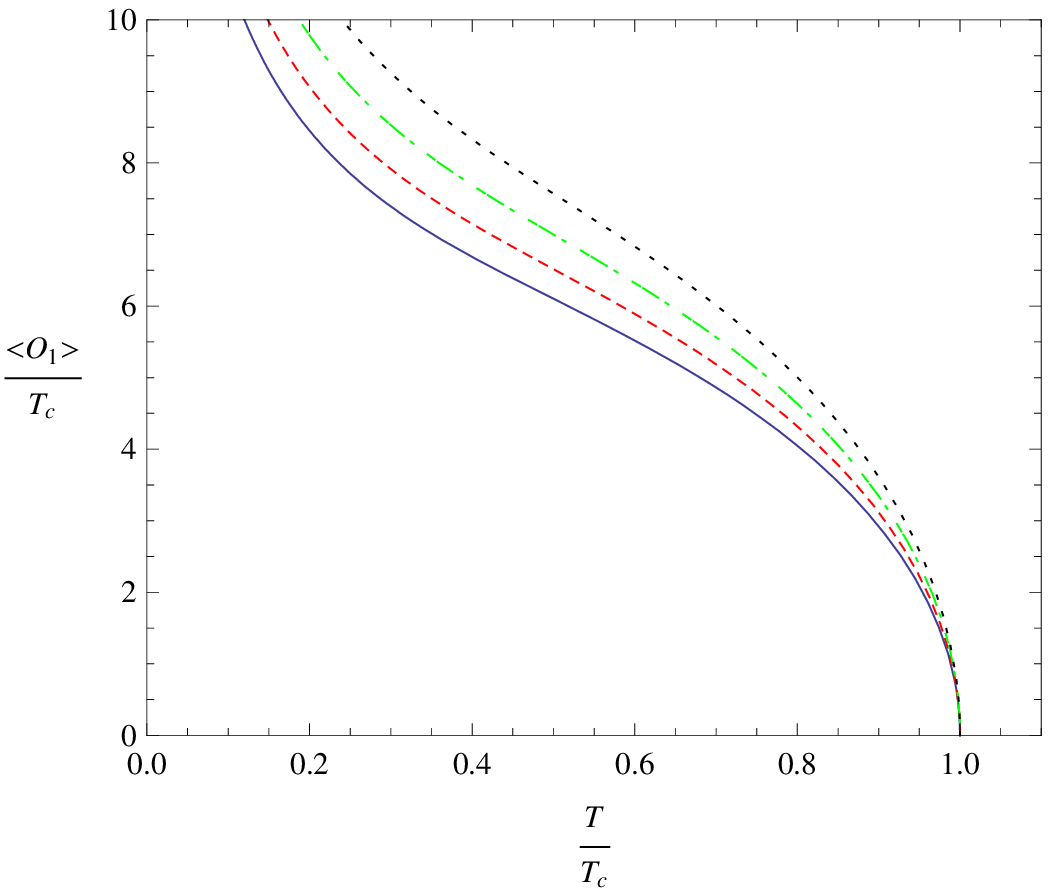}\;\;\;\;
\includegraphics[width=7cm]{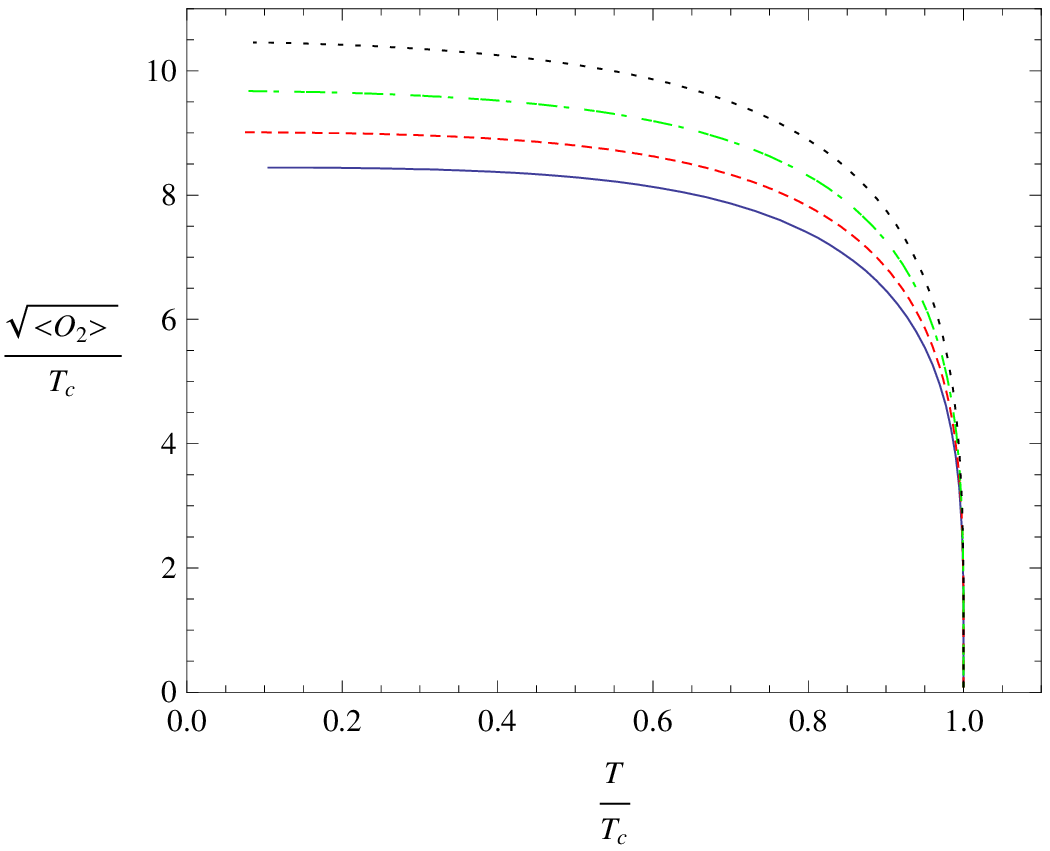}
\caption{The condensates of operators $\mathcal{O}_1$(left) and
$\mathcal{O}_2$(right) versus temperature. The condensates disappear
as $T\rightarrow T_c$. The condensate is a function of temperature
for various values of $b$. Here $b$ is the re-scaled $\tilde{b}$
given by $b=\tilde{b}r^2_H/L^2$. The lowest solid line is $b=0$, the
dashed line is $b=0.25$, the dash-dotted line is for $b=0.5$, and
the top dotted line is $b= 0.75$.}
\end{center}
\end{figure}
\begin{table}[ht]
\begin{tabular}[b]{|c|c|c|c|c|c|}
\hline\hline \;\;\;\;\;\; &
  \multicolumn{2}{c|}{} & \multicolumn{2}{c|}{} \\
  \;\;\;\; $b$\;\;\;\;&
  \multicolumn{2}{c|}{$\mathcal{O}_1$}& \multicolumn{2}{c|}{$\mathcal{O}_2$}\\
 \hline&&&&\\
 0&\;\;$T_c\approx0.226
\rho^{1/2}$&$\langle\mathcal{O}_1\rangle\approx
9.30T_c(1-T/T_c)^{1/2}$&\;$T_c\approx0.118\rho^{1/2}$
&$\langle\mathcal{O}_2\rangle\approx144\;T^2_c(1-T/T_c)^{1/2}$\\
&&&&\\
 \hline &&&&\\
  0.25&$T_c\approx0.222
\rho^{1/2}$&$\langle\mathcal{O}_1\rangle\approx 9.95
  T_c(1-T/T_c)^{1/2}$&$T_c\approx0.111\rho^{1/2}$&
   $\langle\mathcal{O}_2\rangle\approx155.0T^2_c(1-T/T_c)^{1/2}$\\
&&&&\\
 \hline&&&&\\
  0.50&$T_c\approx0.223
\rho^{1/2}$&$\langle\mathcal{O}_1\rangle\approx10.5
  T_c(1-T/T_c)^{1/2}$& $T_c\approx0.103\rho^{1/2}$&$\langle\mathcal{O}_2\rangle\approx174.24T^2_c(1-T/T_c)^{1/2}$\\
&&&&\\
 \hline&&&&\\
  0.75&$T_c\approx0.238
\rho^{1/2}$&$\langle\mathcal{O}_1\rangle\approx 11.4
  T_c(1-T/T_c)^{1/2}$& $T_c\approx0.096\rho^{1/2}$&$\langle\mathcal{O}_2\rangle\approx196.56T^2_c(1-T/T_c)^{1/2}$\\
&&&&\\
 \hline\hline
\end{tabular}
\caption{The critical temperature and the expectation values for the
operators $\mathcal{O}_1$ and $\mathcal{O}_2$ when $T\rightarrow
T_c$ for different values of $b$.}
\end{table}

\begin{figure}[ht]
\begin{center}
\includegraphics[width=6.8cm]{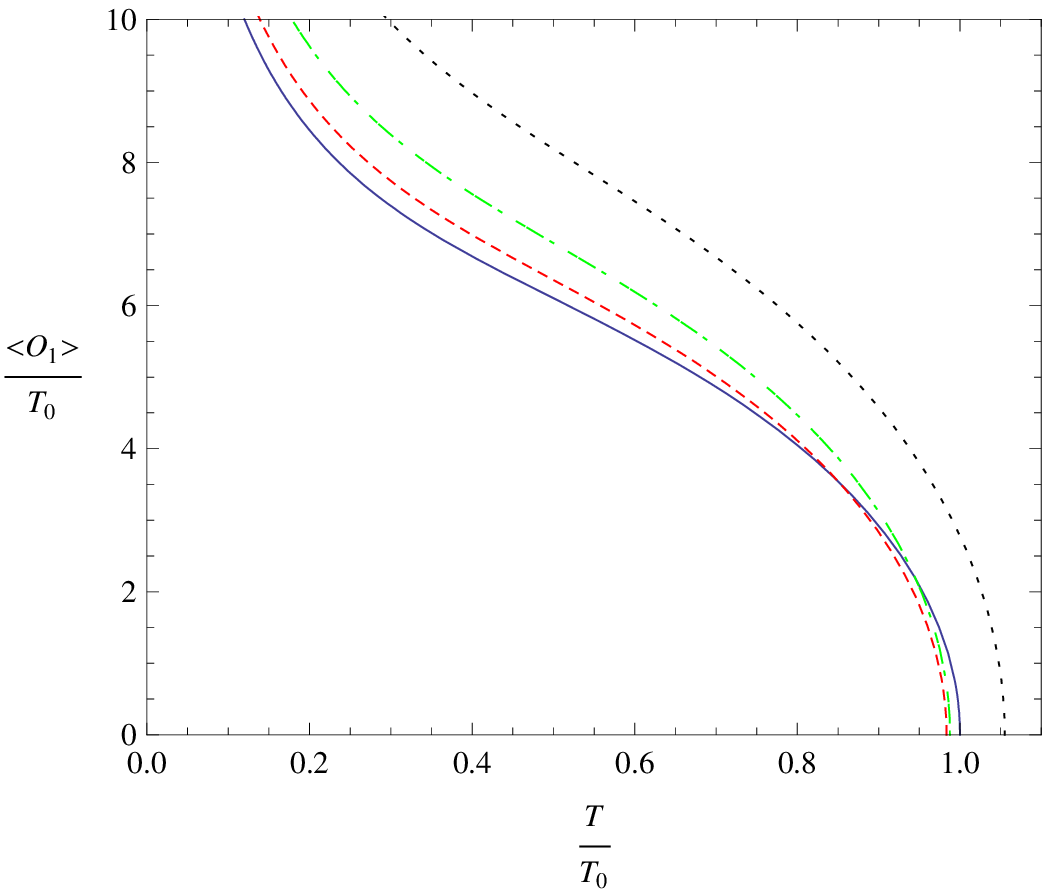}\;\;\;\;
\includegraphics[width=7cm]{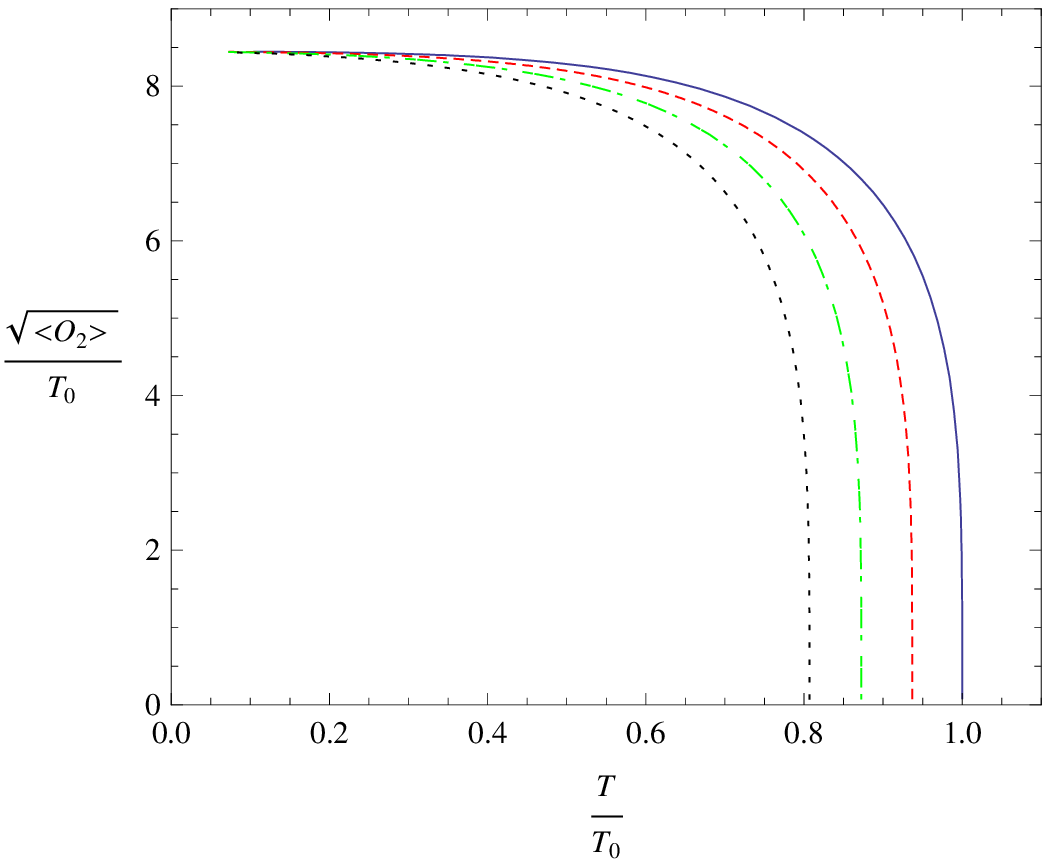}
\caption{The condensates of operators $\mathcal{O}_1$(left) and
$\mathcal{O}_2$(right) versus temperature. The condensate is a
function of temperature for various values of $b$. $T_0$ is the
critical temperature $T_c$ at $b=0$. The solid line is $b=0$, the
dashed line is $b=0.25$, the dash-dotted line is for $b=0.5$, and
the dotted line is $b= 0.75$.  The critical temperature for
$\mathcal{O}_2$ decreases as $b$ increases, while for
$\mathcal{O}_1$ it first decreases and then increases. At very low
temperature, the expectation values $\langle\mathcal{O}_2\rangle$
approach to the same value for different $b$. }
\end{center}
\end{figure}
According to the dual theory, the constants $\mu$ and $\rho$ in the
asymptotic form of $\phi$ are the chemical potential and the charge
density on the boundary, respectively. The expectation values of the
condensate operator $\mathcal{O}$ dual to the scalar field $\psi$.
Since for $\psi$ both of these falloffs are normalizable \cite{ads},
one can impose the boundary condition either $\psi^{(1)}=0$ or
$\psi^{(2)}=0$ to keep the theory stable in the asymptotic AdS
region. Therefore, we have
\begin{eqnarray}
\langle\mathcal{O}_1\rangle=\sqrt{2}\psi^{(1)},\;\;\;\;\;\psi^{(2)}=0,
\end{eqnarray}
or
\begin{eqnarray}
\langle\mathcal{O}_2\rangle=\sqrt{2}\psi^{(2)},\;\;\;\;\;\psi^{(1)}=0.
\end{eqnarray}
As in Refs. \cite{Hs0,Hs01,Hs02,Hs03}, the factor of $\sqrt{2}$  is
a convenient normalization and the index $i$ denotes the scaling
dimension $\lambda$ of its dual operator
$\langle\mathcal{O}_i\rangle$.

In Fig.1 we plot the condensates of operators $\mathcal{O}_1$ and
$\mathcal{O}_2$ for different $b$, which is the re-scaled
$\tilde{b}$ given by $b=\tilde{b}r^2_H/L^2$. We find that if the
temperature $T$ of the black hole  is below the critical temperature
$T_c$, the expectation values of the condensate operators
$\mathcal{O}_1$ and $\mathcal{O}_2$ tend to the constants for fixed
$b$. It is qualitatively similar to that obtain in BCS theory
\cite{bcs}. This means that there exists the holographic
superconductors in the Schwarzschild-AdS black hole with a global
monopole. As in Refs. \cite{Hs0,Hs01,Hs02,Hs03}, the expectation
value $\langle\mathcal{O}_1\rangle$ for fixed $b$ is diverged as
$T\rightarrow 0$, which implies that the backreaction on the bulk
metric can not be ignored in this case. Moreover, we also find that
the expectation values for both the condensate operators increase
with $b$. And then, we fit these condensation curves for the
operators $\mathcal{O}_1$ and $\mathcal{O}_2$ near $T\rightarrow
T_c$ and present the results in the table (I). It is easy to find
that the critical temperature for $\mathcal{O}_2$ decreases as $b$
increases, while for $\mathcal{O}_1$ it first decreases and then
increases. These properties are also shown in Fig.2 in which we
re-plotted the condensation curves for the operators $\mathcal{O}_1$
and $\mathcal{O}_2$ by replacing $T_c$ with $T_0$, where $T_0$ is
the critical temperature at $b=0$. From the Fig.2, it is interesting
for us to find that the expectation values
$\langle\mathcal{O}_2\rangle$ for different $b$ tend to the same
value at the very low temperature. But the expectation values
$\langle\mathcal{O}_1\rangle$ do not possess this behavior. These
differences could be explained by the fact that the operator
$\mathcal{O}_2$ corresponds to a pair of fermions, while the
operator $\mathcal{O}_1$ to a pair of bosons \cite{Hs0}.

\section{The electrical conductivity}

In this section we will compute the electrical conductivity by
perturbing the Maxwell field. For simplicity, we adopt to the probe
approximation and neglect the effect of the perturbation of
background metric. Assuming that the perturbation of the vector
potential has a form $\delta A_x =A_x(r)e^{-i\omega t}$ \cite{Hs0},
we can obtain the linearized equation of motion
\begin{eqnarray}
A_x^{''}+\frac{f'}{f}A_x'
+\bigg(\frac{\omega^2}{f^2}-\frac{2q^2\psi^2}{f}\bigg)A_x=0.\label{de}
\end{eqnarray}
Since there exists only the ingoing wave at the black hole horizon,
the boundary condition on $A_x$ near $r\sim r_H$ in the
Schwarzschild-AdS black hole spacetime with a global monopole can be
expressed as
\begin{eqnarray}
A_x= f^{-\frac{i\omega r_H}{(3r^2_H-\tilde{b})}}~.
\end{eqnarray}
From Eqs. (\ref{b1}) and (\ref{de}), it is easy to obtain that at
the spatial infinity $A_x$ possesses the asymptotic behavior
\begin{eqnarray}
A_x=A^{(0)}_x+\frac{A^{(1)}_x}{r}+....
\end{eqnarray}
\begin{figure}[ht]
\begin{center}
\includegraphics[width=7cm]{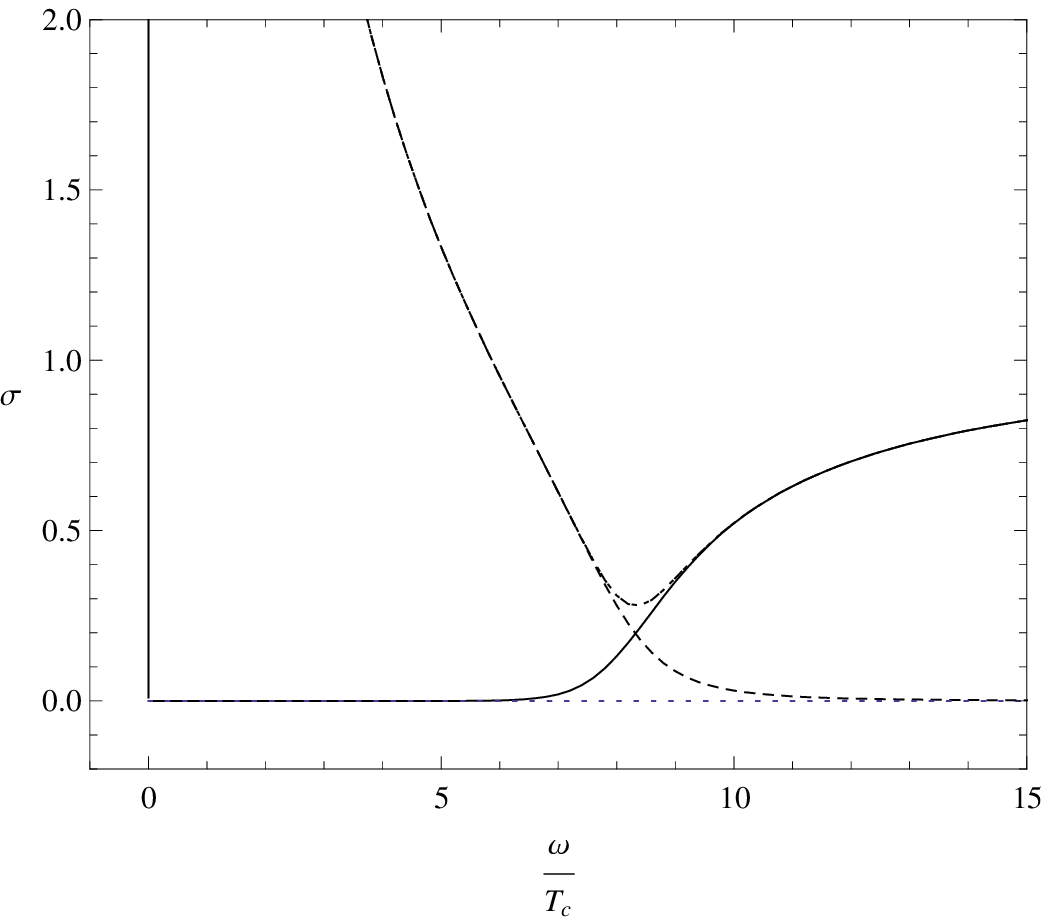}\;\;\;\;
\includegraphics[width=7cm]{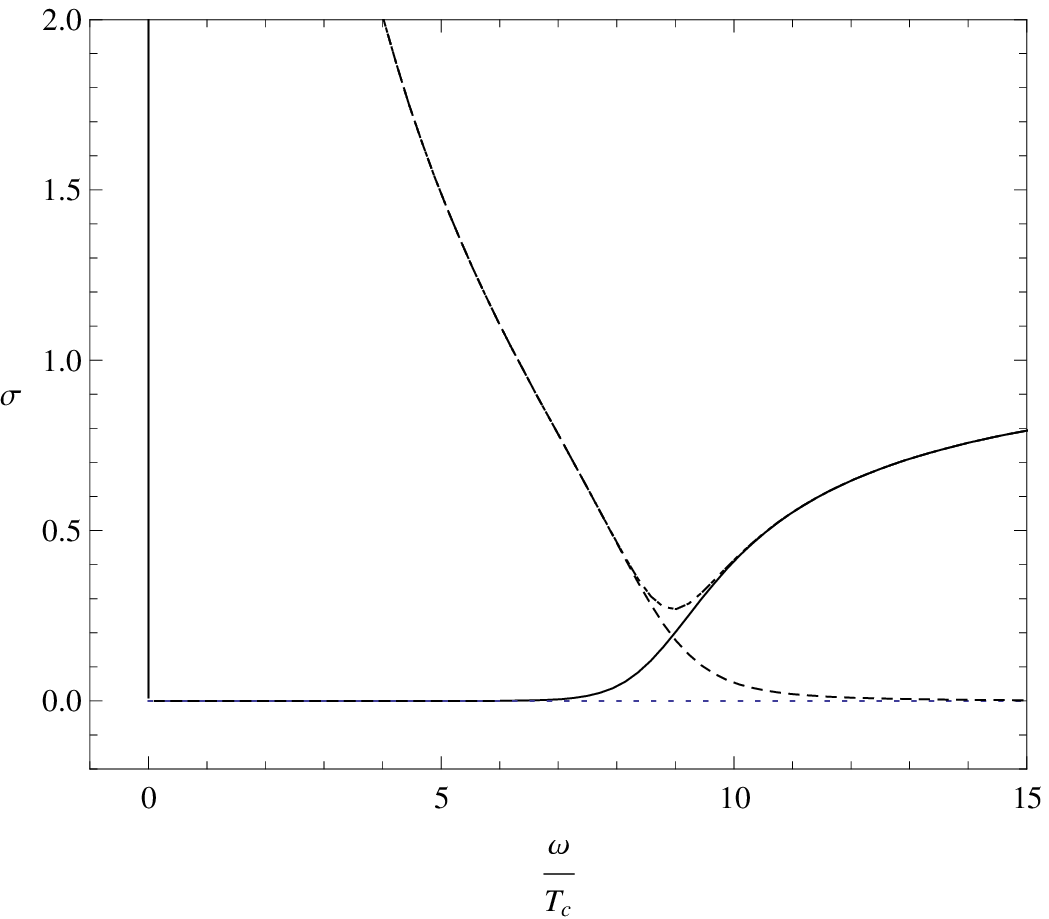}\\
\includegraphics[width=7cm]{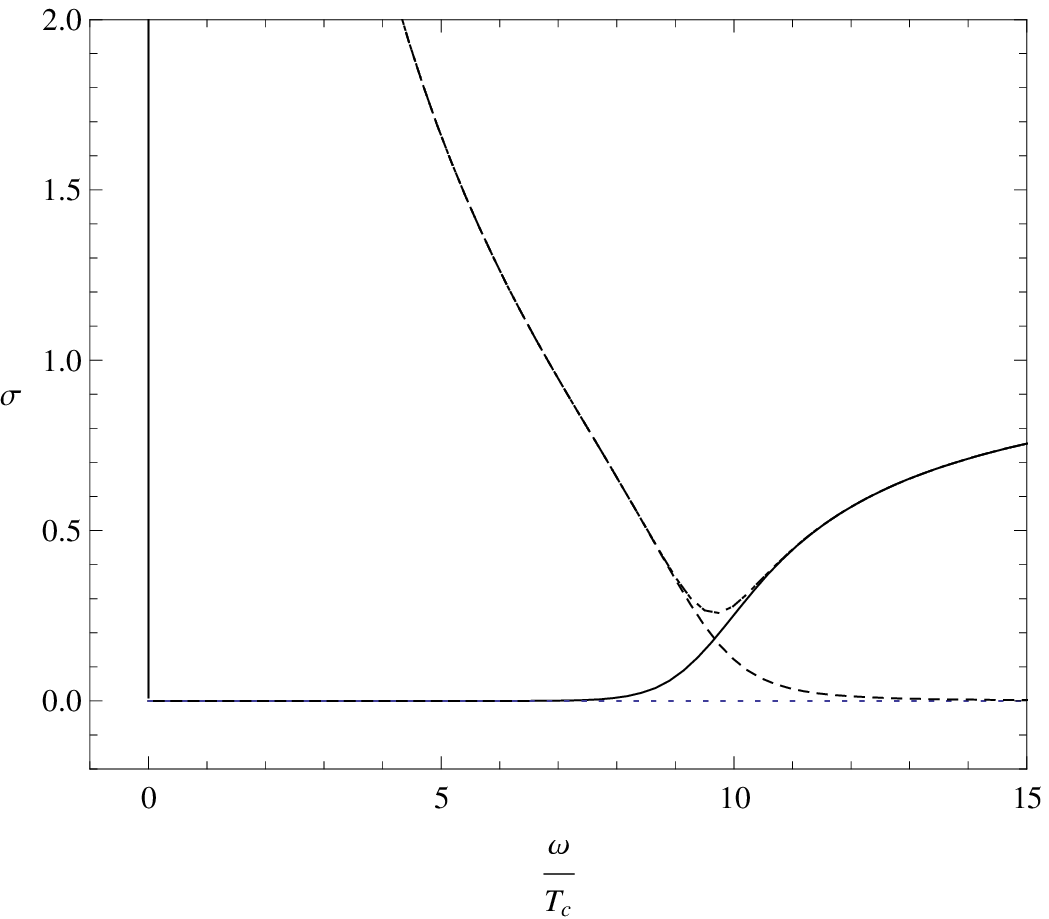}\;\;\;\;\includegraphics[width=7cm]{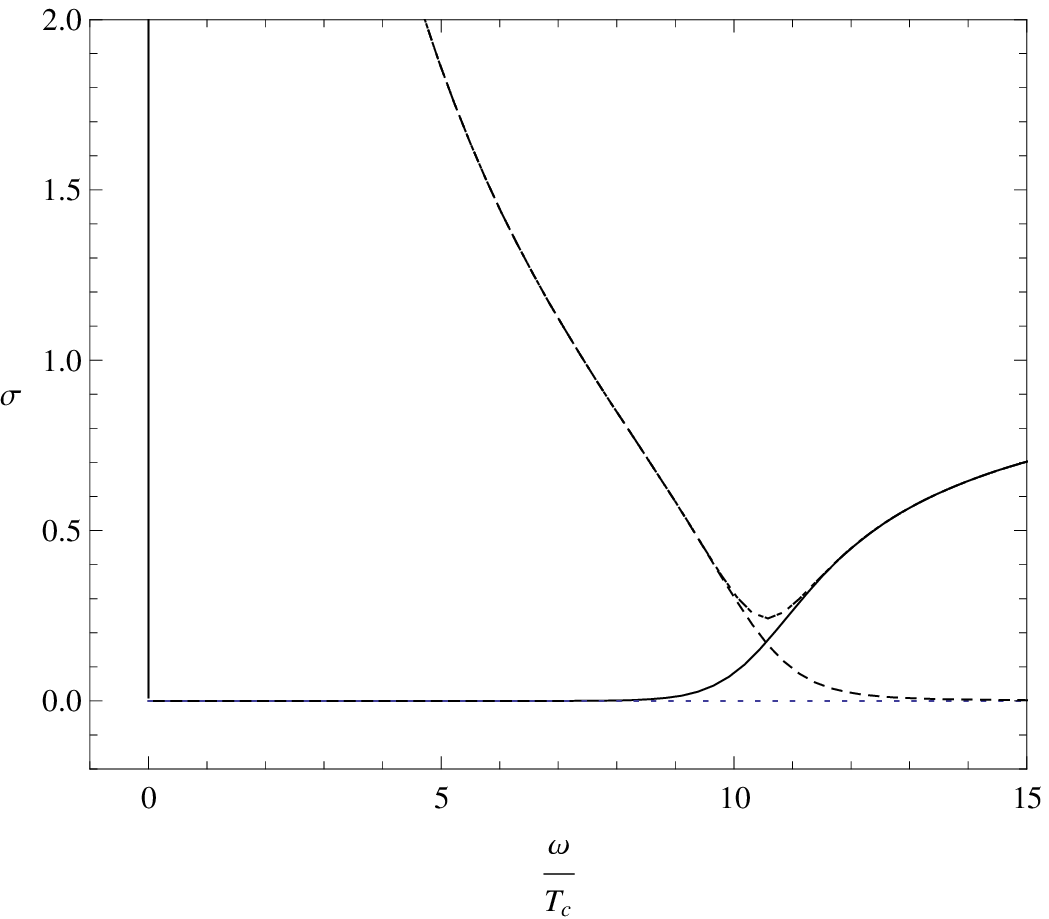}
\caption{The frequency dependent conductivity with different values
of $b$ for the operator $\langle\mathcal{O}_1\rangle$. Each plot is
at low temperatures, about $T/Tc\approx0.20$. The up-left, up-right,
bottom-left and bottom-right are corresponding to the cases $b=0$,
$0.25$, $0.5$ and $b=0.75$, respectively. The solid, dashed and
dash-dotted curves denote the real part $Re \sigma$, the imaginary
part $Im \sigma $ and the module $|\sigma|$ of conductivity,
respectively. }
\end{center}
\end{figure}
From the AdS/CFT, it is well known that $A^{(0)}_x$ and $A^{(1)}_x$
in the bulk corresponds to the source and the expectation value for
the current on the CFT boundary, respectively. Thus, we have
\cite{Hs0}
\begin{eqnarray}
A_x=A^{(0)}_x,\;\;\;\;\;\;\langle J_x\rangle=A^{(1)}_x.
\end{eqnarray}
According to the Ohm's law, we can calculate the conductivity by
\begin{eqnarray}
\sigma(\omega)=\frac{\langle J_x\rangle}{E_x}=-\frac{i\langle
J_x\rangle}{\omega A_x}=-i\frac{A^{(1)}_x}{\omega A^{(0)}_x}.
\end{eqnarray}
\begin{figure}[ht]
\begin{center}
\includegraphics[width=7cm]{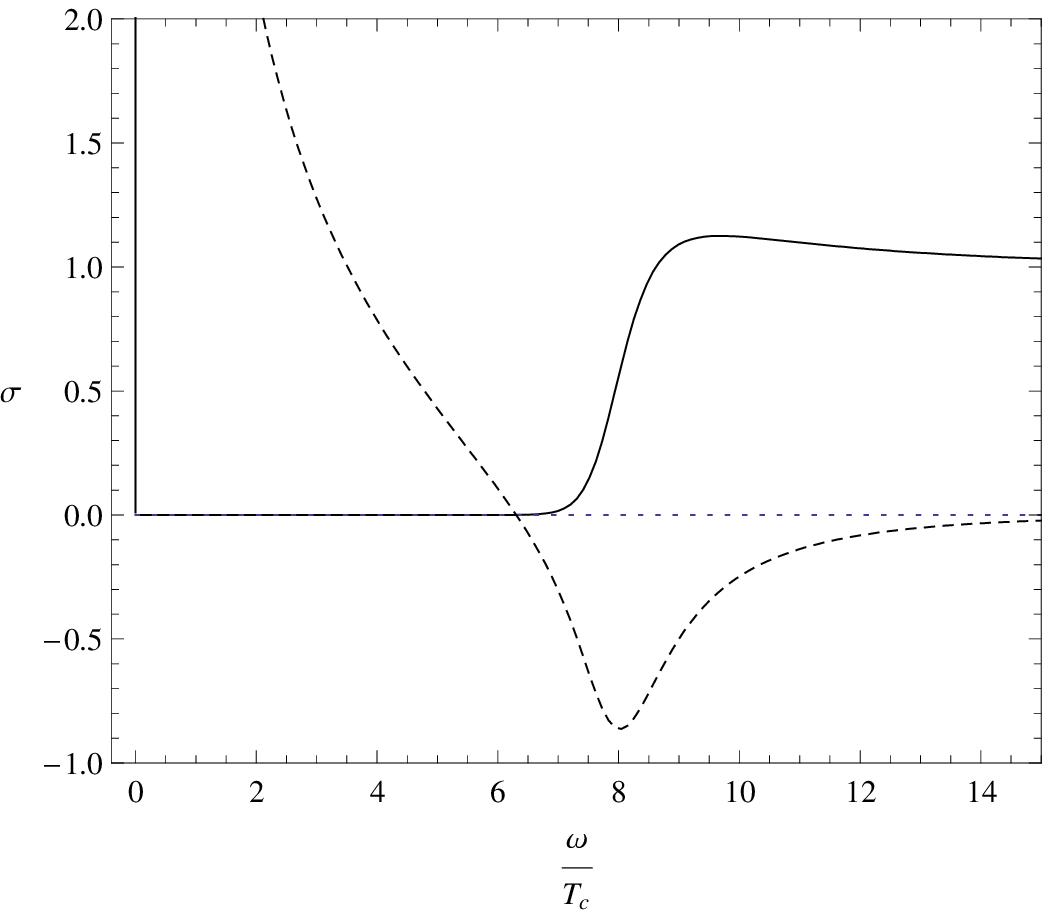}\;\;\;\;
\includegraphics[width=7cm]{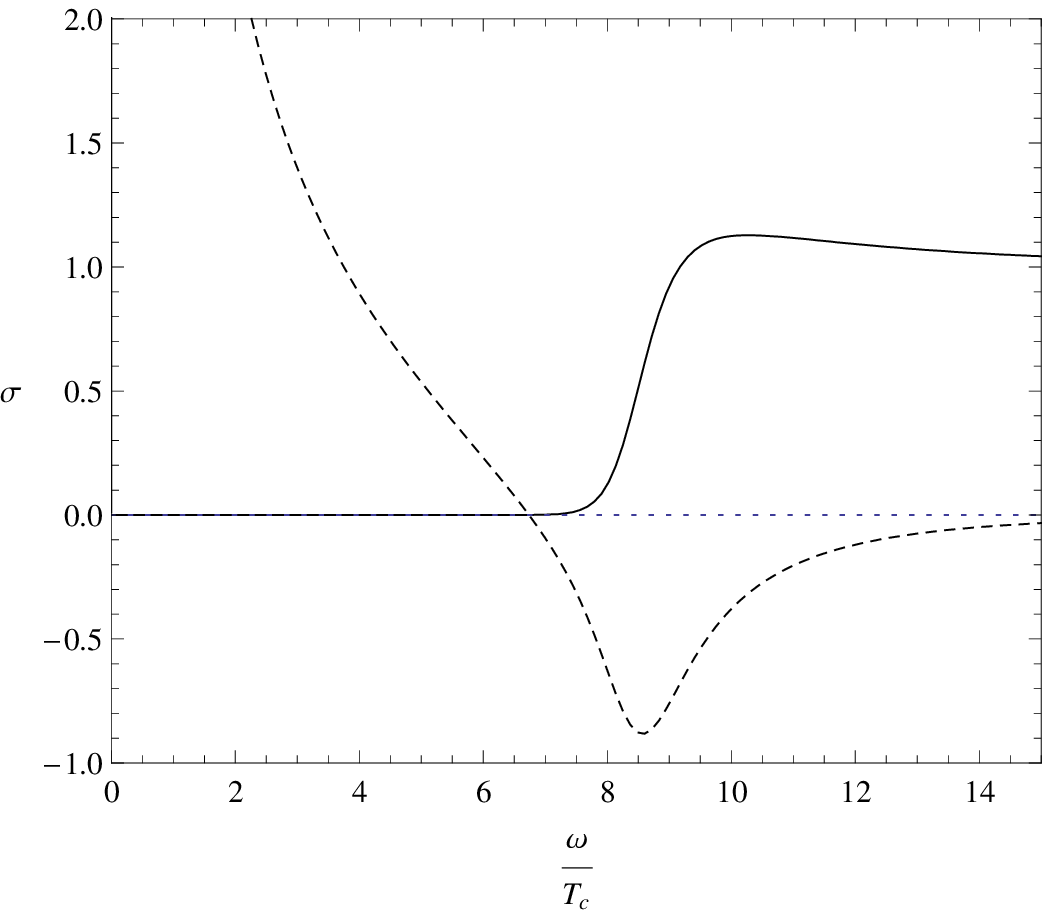}\\
\includegraphics[width=7cm]{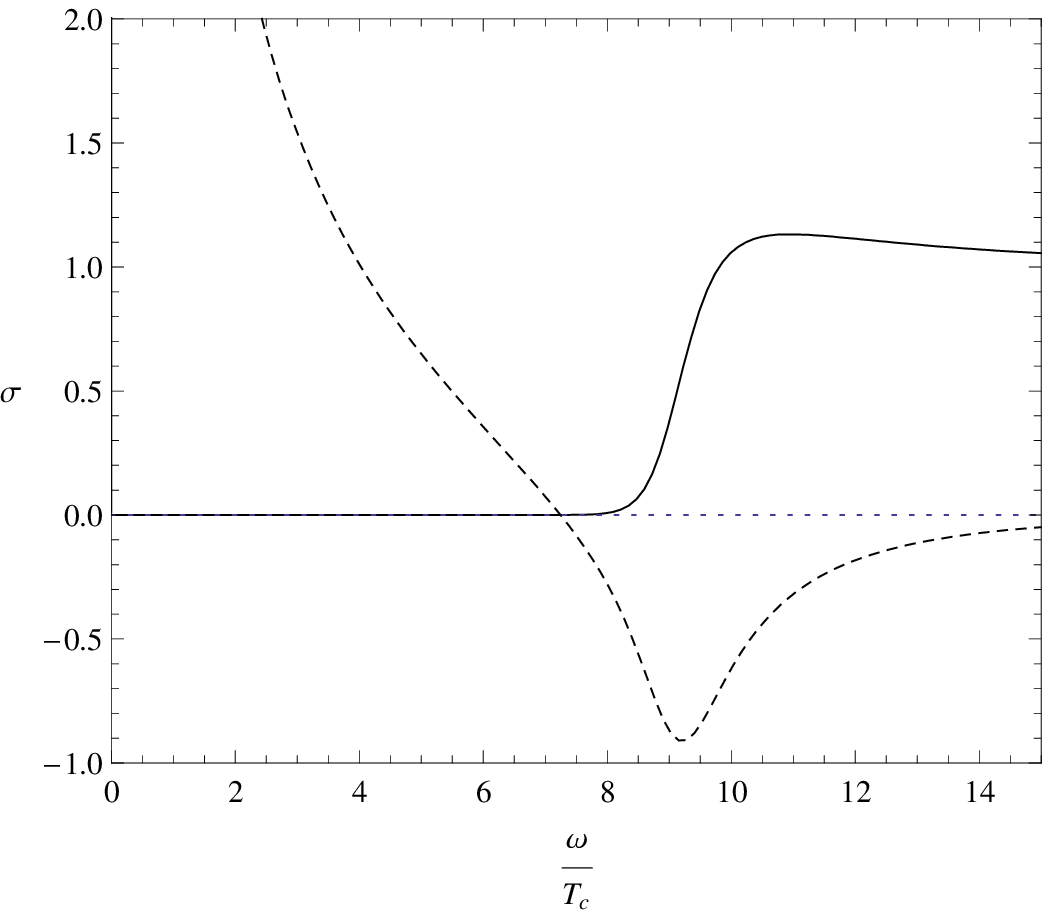}\;\;\;\;
\includegraphics[width=7cm]{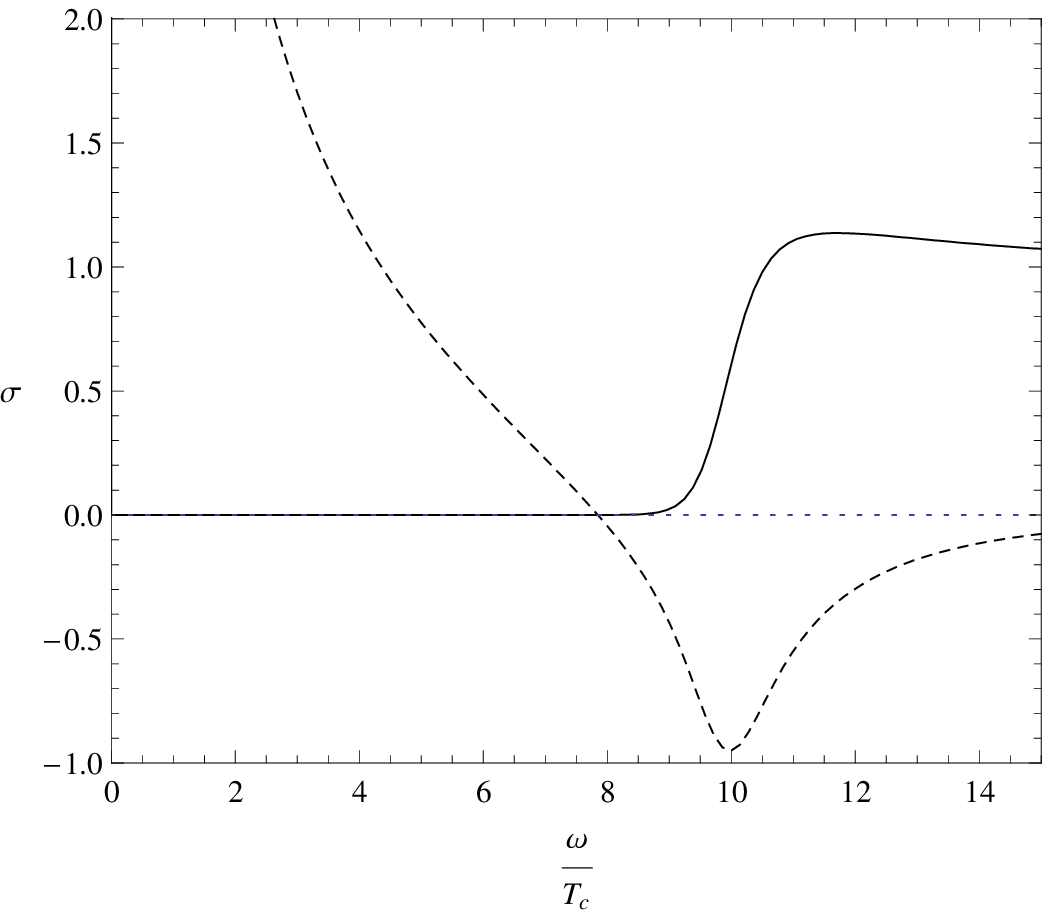}
\caption{The frequency dependent conductivity with different values
of $b$ for the operator $\langle\mathcal{O}_2\rangle$. Each plot is
at low temperatures, about $T/Tc\approx0.10$. The up-left, up-right,
bottom-left and bottom-right are corresponding to the cases $b=0$,
$0.25$, $0.5$ and $0.75$, respectively. The solid and dashed curves
denote the real and imaginary parts of conductivity, respectively. }
\end{center}
\end{figure}
In Figs. 3 and 4, we plot the frequency dependent conductivity with
different $b$ for the operators  $\mathcal{O}_1$ at temperature
$T/Tc\approx0.20$ and $\mathcal{O}_2$ at $T/Tc\approx0.10$,
respectively. For the condensate operator $\mathcal{O}_1$, we have
$\lambda=1$, \textit{i.e.}, $m^2L^2=-2$, which is above the
Breitenlohner-Freedman bound \cite{BF}. We find that the real part
increases and imaginary part of the conductivity decreases
monotonously with the frequency $\omega$ for fixed $b$. Each modules
of $\sigma$ has a minimum value, which is the similar to that
obtained in Ref. \cite{GT}. Moreover, from the figure (3), we also
find that the value of $\frac{\omega_g}{T_c}$ increases with the
increase of $b$. For the condensate operator $\mathcal{O}_2$
($\lambda=2$), we find from the figure (4) that, as the frequency
$\omega$ increases for fixed $b$, the real part possesses the
similar behavior as that in the case $\lambda=1$. However, the
imaginary part has a minimum value in this case and the position of
the minimum value moves along right when $b$ increases . Moreover,
as in other literatures \cite{Hs0,Hs01,a0,a40}, we find that at
$\omega=0$ the real part of conductivity  behaves as a delta
function and the imaginary part exists a pole in the background with
the global monopole. Similarly, this can be explained by using the
Kramers-Kronig relation.

\section{Summary}

In this paper we studied the holographic superconductors in the
Schwarzschild-AdS black-hole spacetime with a global monopole
through a charged complex scalar field. We adopted to the probe
approximation and solved the coupled nonlinear equations of the
system for different $b$, which is related to the symmetry breaking
scale $\eta$ due to the presence of the global monopole. We found
that the global monopole presents  the different effects on the
different condensates $\mathcal{O}_1$ and $\mathcal{O}_2$. As $b$
increases, the critical temperature for $\mathcal{O}_2$ decreases,
while for $\mathcal{O}_1$ it first decreases and then increases.
Moreover, we also find that the expectation values
$\langle\mathcal{O}_2\rangle$ for different $b$ tend to the same
value at very low temperature, while the expectation values for
$\langle\mathcal{O}_1\rangle$ do not possess this behavior. These
differences could be explained by the fact that the operator
$\mathcal{O}_2$ corresponds to a pair of fermions and the operator
$\mathcal{O}_1$ to a pair of bosons \cite{Hs0}. Moreover, we
discussed the electric conductive at low temperature. Our results
showed that the real part of the conductivity increases as the
frequency $\omega$ increases for all $b$. For the case $\lambda=1$
the imaginary part decreases monotonically with $\omega$ and the
module of $\sigma$ has a minimum value. The position of the minimum
value is near $\frac{\omega_g}{T_c}\approx8$ and moves along right
with $b$. For $\lambda=2$, the position of the minimum value of the
imaginary part is similar to the position of the minimum value of
module in the case $\lambda=1$. These results can help us know more
about holographic superconductors in the asymptotic AdS black holes.

\begin{acknowledgments}
We thank Professor Bin Wang and Dr Qiyuan Pan for their helpful
discussions and suggestions. This work was partially supported by
the National Natural Science Foundation of China under Grant
No.10875041 and the construct program of key disciplines in Hunan
Province. J. L. Jing's work was partially supported by the National
Natural Science Foundation of China under Grant No.10675045,
No.10875040 and No.10935013; 973 Program Grant No. 2010CB833004 and
the Hunan Provincial Natural Science Foundation of China under Grant
No.08JJ3010.
\end{acknowledgments}

\end{document}